# NEW Fe I LEVEL ENERGIES AND LINE IDENTIFICATIONS FROM STELLAR SPECTRA. III. INITIAL RESULTS FROM UV, OPTICAL, AND INFRARED SPECTRA

Ruth C. Peterson
SETI Institute, 339 Bernardo Ave., Suite 200, Mountain View, CA 94043
e-mail: peterson@ucolick.org

Robert L. Kurucz
Center for Astrophysics | Harvard & Smithsonian, 60 Garden Street, Cambridge, MA 02138

*Short title:* New Fe I Energies and Line Identifications Incorporating Stellar Infrared Spectra

**ABSTRACT**

The spectrum of neutral iron is critical to astrophysics, yet furnace laboratory experiments cannot reach high-lying Fe I levels. Instead, Peterson & Kurucz and Peterson et al. adopted ultraviolet (UV) and optical spectra of warm stars to identify and assign energies for 124 Fe I levels with 1900 detectable Fe I lines, and to derive astrophysical gf values for over a thousand of these. An energy value was assumed for each unknown Fe I level, and confirmed if the wavelengths predicted in updated Kurucz Fe I calculations matched the wavelengths of four or more unidentified lines in the observed spectra. Nearly all these identifications were for LS levels, those characterized by spin-orbit coupling, whose lines fall primarily at UV and optical wavelengths. This work contributes nearly a hundred new Fe I level identifications. Thirty-nine LS levels are identified largely by incorporating published positions of unidentified laboratory Fe I lines with wavelengths <2000 Å. Adding infrared (IR) spectra provided 60 Fe I jK levels, where a single outer electron orbits a compact core. Their weak IR lines are searchable, because their mutual energies obey tight relationships. For each new Fe I level, this work again makes publicly available its identification, its energy, and a list of its potentially detectable lines with theoretical gf values, totaling >16,000 lines. For over two thousand of these, this work provides astrophysical gf values adjusted semiempirically to fit the stellar spectra. The potential impact of this work on modeling UV and IR stellar spectra is noted.

## 1. OVERVIEW

This work continues investigations by Peterson & Kurucz (2015; hereafter PK15) and Peterson et al. (2017; hereafter PKA17) to derive new identifications and energies for Fe I levels that are too highly excited to have been determined in earlier laboratory and solar analyses. These two works adopted a wide range of stellar spectra observed in the ultraviolet (UV) and optical wavelengths as the "laboratory" source to seek levels whose lines were strong and isolated enough to be distinctly discerned. Matching four lines predicted for each level to observed wavelengths with a single energy shift determined the new level energy and its identification. In this work, incorporating laboratory positions for unidentified lines with wavelengths <2000 Å and adding infrared (IR) spectra of the Sun and the cool giant Arcturus has enabled the determination of about a hundred new Fe I levels. Their energies, identifications, and associated Fe I lines potentially detectable in solar-type dwarfs and moderately cooler giants of solar metallicity and higher, including astrophysical gf values where feasible, are presented here and made publicly available.

Section 2 summarizes the importance of extending Fe I level identifications, and the procedure and the results of previous work. Section 3 describes the identification of 39 new levels, found primarily by adding the laboratory positions of unidentified lines at wavelengths blueward of the useful range of stellar UV spectra. Section 4 outlines how incorporating IR spectra can yield identifications for Fe I levels of a different structure whose lines fall primarily in the IR. Section 5 presents the IR spectra adopted, and Section 6 discusses the calculation of IR spectra to match them. Section 7 describes modifications to the search procedure adopted in the IR. Section 8 evaluates and discusses uncertainties. Section 9 summarizes the results, and Section 10 describes the status, prospects, and impact of the identification of Fe I levels.

## 2. PREVIOUS WORK

Astrophysical observations are dramatically improving. New cosmic surveys are recording a plethora of ever-fainter sources. Automated analysis offers the promise of rapid interpretation. Often lacking, however, are the laboratory astrophysics data needed to analyze the new observations. These are the basic atomic and molecular parameters necessary to model at all wavelengths, from UV to IR, the spectral absorption and emission of supernovae, nebulae, and the hot outer layers of stars and their galaxies, at both low and high redshifts.

Especially critical are line parameters for the neutral iron atom, due to its high cosmic abundance and its complex line spectrum. In the Sun, the Fe I lines of neutral iron have long been known to dominate both the optical region (e.g., Moore et al. 1966) and the UV (e.g., Tousey 1988). While energy levels, wavelengths, and transition probabilities (gf values) can often be derived theoretically for light, simple atoms, the Fe I spectrum is best characterized observationally. From a laboratory furnace, Brown et al. (1988; hereafter B+88) provided determinations of wavelengths of Fe I spectral absorption in the UV and optical regions. Nave & Johansson (1993a, 1993b) then provided Fe I level energies up to ~60,800 cm$^{-1}$ with uncertainties <0.005 cm$^{-1}$, and wavelengths for over 2000 Fe I lines from 1700 Å to 5 μm. They noted that a myriad of lines of high-lying unidentified Fe I levels still remained in the solar spectrum.

In the ~2000 K laboratory furnace, the population of highly excited Fe I levels is severely depleted by the exponential Boltzmann temperature dependence. In contrast, stars at ~6000 K like the Sun are warm enough to significantly populate these levels without overly ionizing Fe I to Fe II. In the spectra of such warm stars, the resulting multitude of unidentified Fe I lines leads to poor spectral modeling, notably a serious underestimate of observed UV flux distributions by calculations that lack them (PK15, Figure 1).

Fe I level energies can be predicted through quantum-mechanical calculations. For example, Kurucz (2011, 2017, 2018) has run least-square fits based on the Cowan (1981) codes to predict whole Fe I transition arrays and the resulting wavelengths and gf values of the lines of each level across all wavelengths. However, these are not nearly as accurate as laboratory derivations, due to the complexity of the Fe I atom and mixing in the eigenvectors that characterize each level. Owing to the sizable errors in theoretical level energies, adopting these theoretical wavelengths leads to typical errors of 10 Å in the UV and 100 Å in the IR.

The predicted line parameters suffice for modeling low-resolution flux distributions of stars across a wide temperature range, but are woefully inadequate where discernment of lines of other species is essential. This is critical for many research areas, such as two recently reviewed by Matteucci (2021): the assembly of the Galaxy following galactic nucleosynthesis and star formation at the earliest epochs (e.g., Sneden et al. 2008), and the corresponding origin and development of the Galactic bulge (Ryde et al. 2016; Barbuy et al. 2018). Unrecognized absorption by unidentified Fe I lines may lead to abundance overestimates for abundances of critical trace elements whose lines occur only in the UV (Peterson et al. 2020; Peterson 2021). As discussed in Section 10, their presence affects the placement of the IR continuum in strong-lined giants such as those of the Galactic bulge (e.g., Ryde et al. 2016, Figure 3). Identifying unknown Fe I lines is thus critical to facilitate the total scientific harvest from existing high-resolution spectrographs, even those with resolutions ~50,000 such as Keck HIRES (Vogt et al. 1994) and CRIRES (Käufli et al. 2004), and from those under development such as the ELT-HIRES spectrograph (Maiolino et al. 2013).

Consequently, PK15 and PKA17 adopted UV and optical spectra of warm stars as the "laboratory source." Extending the approach of Castelli & Kurucz (2010) to identify Fe II levels in a single hot B star, these authors compared the observed optical and UV spectra of a variety of warm and cool stars with theoretical synthetic spectral calculations generated with the Kurucz (1993) program SYNTHE that were run for each individual star. The atmospheric models for these calculations were interpolated in the model grid of Castelli & Kurucz (2003) to the parameters given in Table 1 of PK15. The line parameters input to these calculations were constructed from the Kurucz (2011) list for known (but not predicted) Fe I lines, along with Kurucz line lists for other species, both atomic and molecular. To better reproduce the observed line strengths, atomic gf values in the original lists were revised line-by-line, while those for molecules were altered by band.

The identification of each particular unknown level depended on the Kurucz (2011) Fe I quantum-mechanical calculations, based on the precepts of Cowan (1981). An energy was simply assumed for an unknown level; it was confirmed if the resulting adjusted positions of at least four of its associated predicted spectral lines matched the exact wavelengths of at least four unidentified lines in all relevant stellar spectra. Furthermore, a single adopted gf value for each such line, which was usually modestly lower than the predicted value, had to provide a calculated strength that matched approximately the observed strength of that line in each stellar spectrum. As Section 1 of PKA17 emphasizes, the observed spectra deliberately spanned a wide range in both line strength and stellar temperature; matching all their strengths simultaneously provided a more stringent test than matching a few similar spectra. Once four such line matches had confirmed the new identifications of about 20 Fe I levels, the Kurucz Fe I predicted calculations were run again. This significantly improved the predicted energies, wavelengths, and gf values for the next round of Fe I searches. Following this procedure, these two publications provided identifications and energies for a total of 124 previously unidentified Fe I levels, resulting in nearly 3000 detectable lines in the UV/optical and a similar number of lines likely detectable in the IR.

## 3. LEVELS WHOSE LINES FALL PRIMARILY IN THE UV AND OPTICAL

Nearly all the PK15 and PKA17 Fe I identifications were for LS levels, whose structure is characterized by spin-orbit coupling. Lines of LS levels are strongest in the UV and optical, so the preceding works targeted levels with strong lines in relatively unblended sections of these regions. The remaining unidentified levels tend to have weaker lines, or strong lines that fall in heavily obscured UV regions. Two of the latter regions are at wavelengths below 2000 Å, and throughout the 2600 – 2700 Å region shown in Figure 1 of PK15. Peterson et al. (2020, Section 5) demonstrated that the former region is heavily blanketed by unidentified Fe I lines of high-excitation levels of odd parity, whose lower even-parity levels lie near the ground state, and that their even-parity counterparts populate the latter region, since the lowest odd-parity Fe I levels fall well above the ground state.

For this work, an additional HST STIS UV spectrum covering 1630 – 1900 Å was obtained for the warm metal-poor star HD 84937. While this enabled the detection of strong unidentified lines at wavelengths <1900 Å, the declining spectral quality and increased blending at these wavelengths limit the success of searches for further Fe I level identifications (PKA17, Section 6). Figures 1—3 of Peterson et al. (2020) illustrate directly how the crowding of unidentified lines coupled with the low signal-to-noise ratio (S/N) in the stellar spectra make line identification difficult. However, due to the low excitation levels of Fe I lines in this region, the laboratory furnace provides a valuable resource. The present work has taken advantage of the laboratory Fe I line positions in this region, especially those for which B+88 could not make a definitive identification, but could provide a lower and upper energy and J value, based on the spacing in wavelength at successively increasing lower levels of excitation.

Incorporating these B+88 lines and levels provided identifications for nearly three dozen Fe I levels that otherwise had too few UV and optical stellar spectral line matches. To identify fully a B+88 level assigned an energy E and angular momentum J, a predicted level was selected of the same J and similar E whose UV line gf values, when reduced to reflect the population of the lower level at the 2000 K furnace temperature, provided consistency with the B+88 line-intensity measurements. The level energy was assigned accordingly; agreement to 0.02 cm$^{-1}$ to the B+88 line positions was demanded in the shifted predicted line positions. Each such match was counted as a line detection. Levels with only three detections were accepted when no other LS level with the same J and a predicted energy within ±150 cm$^{-1}$ remained unassigned. In addition, several levels with at least one line detected at longer wavelengths were identified by similarly matching B+88 lines for which no J nor energy was provided.

Thirty-three of the 39 new LS level identifications were made in this way, all for levels of odd parity. The largest difference between the predicted and adopted energy of a new LS level was 269.8 cm$^{-1}$. For all others this difference was below 200 cm$^{-1}$, and exceeded 150 cm$^{-1}$ for only three of these. Ambiguities in assigning levels still exist, especially at higher energies. As an example, the two LS levels with the highest energies may have been swapped, as their vector structures and line gf values are similar, and the predicted energies are not yet constrained tightly enough to fully distinguish them.

The remaining six new LS identifications relied as before on stellar lines present across the entire spectrum. Three met the PK15 and PKA17 criteria directly, including the two with the highest energies. The PK15 odd levels 3.0 6p 5G5D3D and 4.0 (4F)4p 5D were removed as their energies were assigned to the 3.0 4s6D7p 7F and the 4.0 4s6D8p 7F levels, providing a better match. Three other levels with J > 6 were accepted but with only two matches, since each match was made for an unusually strong line with a high gf value. All but one of these six newly identified LS levels are also of odd parity.

Among levels of even parity, similar checks of LS levels with large differences between observed and predicted energies led to the rejection of one level and the reassignment of the energy of two levels identified by PKA17. The difference was 265 cm$^{-1}$ for the rejected PKA17 LS level, 3.0 4s6D8s 5D, which had six lines detected. It will be reinstated if no other assignment can be made to its energy. One even PK15 level was also dropped, and two revised; and one odd PKA17 level was revised. Applying the procedures for odd jK levels described below in Section 7 necessitated changes for published jK levels as well. Table 1 lists these revisions.

## 4. LEVELS WITH LINES IN INFRARED STELLAR SPECTRA

In solar-type stars, almost all unidentified UV/optical Fe I lines arise from a transition to an unknown upper level from a known low-lying level. The line strength depends on the population of the low level, which is only modestly diminished at these temperatures by the exponential energy dependence of the Boltzmann factor. However, unidentified Fe I lines also appear in the IR, beyond 1μm. These arise from two high levels, one or both of which may be unknown. In either case, these unidentified Fe I IR lines are rather weak. The absorbing population is diminished by the Boltzmann factor of the lower level, which is usually at high energy since the vast majority of Fe I levels at lower energy are known. Consequently, unidentified lines in IR spectra of stars of solar temperature and cooler are much weaker than the unidentified UV and optical lines studied previously. IR searches thus demand higher accuracy, in both the observational spectra and in the theoretical spectral calculations.

Including IR spectra nonetheless assists such searches, as discussed and illustrated in Section 5 and Figure 1 of PKA17. IR spectra not only expand the number of potential line detections for a given level, but also provide a more accurate determination of the level energy. The latter arises because the spectral line profile in the relevant wavenumber (energy) space narrows steadily to the red: the stellar line profile is constant in velocity space, and so varies as $\delta\lambda/\lambda$.

Furthermore, IR spectra enable the determination of Fe I levels with a different atomic structure. The newly identified levels described in Section 3, and virtually all the levels identified by PK15 and PKA17, have LS coupling, the spin-orbit interaction between the outer and innermost electrons of neutral iron. Incorporating IR stellar spectra in this work, especially those of the Sun, has opened up the determination of energies for levels represented by jK coupling, in which a core configuration is accompanied by a single, relatively isolated electron. Such levels rarely have distinct UV lines, but have a myriad of IR lines, the vast majority of which are weak.

## 5. SELECTION OF INFRARED STELLAR SPECTRA

The very-high-quality IR spectra of the Sun provided the greatest number of jK line detections. IR spectra of two cool giants, especially that of the mildly metal-poor giant Arcturus, were also consulted to distinguish whether a particular unidentified feature is indeed due to Fe I, and to sort out contamination by other low-excitation lines.

The limited number of IR stellar spectra searched arose from three aspects pertinent to IR but not UV-optical Fe I searches. (a) In the IR, unidentified Fe I lines become very weak, for the population of the lower level is reduced by its substantially higher excitation (Section 4). Consequently, high S/N and spectral resolution are especially crucial. By these criteria, the Sun is by far the best-observed star in the IR. (b) IR spectra obtained from the ground are strongly affected by absorption by the Earth's atmosphere, which appears sporadically across a wide range of IR wavelengths and blocks some broad regions completely (Hinkle et al. 1995, Table 3). Space-based spectra of the Sun were adopted where available; elsewhere, ground-based spectra for the Sun and Arcturus were adopted that attempted to correct for this absorption. (c) Strong molecular absorption occurs throughout the IR in spectra of cool giants such as Arcturus (Hinkle et al. 1995, Table 5), and is much weaker but nonetheless present in the solar IR spectrum as well. The Arcturus spectrum provides a useful comparison to judge whether a given absorption feature in the solar spectrum is blended (or obliterated) by molecular absorption. The spectrum PKA17 adopted for the red horizontal branch star HD 95870 was also included where it was available. However, spectra of the cooler, metal-rich giant µ Leo (HD 85503) were not used because of strong molecular contamination, notably that of CN.

A comparison of the spectra of the Sun and Arcturus was very useful to reveal the atomic species giving rise to a particular absorption feature or blend. Mildly metal-poor, Arcturus has high abundances with respect to iron of several atomic species such as Mg I, Si I, and Ca I (e.g., Peterson, et al. 1993), whose lines are significantly present in the IR (Hinkle et al. 1995, Table 4).  Fig 1(b) of PKA17 illustrates that, while the Fe I line at 1776.5234 nm is about the same strength in the spectra of the Sun and Arcturus, the Mg I lines near 1774.96, 1775.4, and 1776.2 nm all become much stronger in Arcturus than in the Sun. In contrast, HD 95870 has near-solar abundances (Afşar et al. 2018).

For the Sun, for wavelengths beyond 2.1µm, telluric-free Fourier transform spectroscopy (FTS) spectra are available from two space missions designed to secure high-resolution spectra of the Earth's atmosphere: ATMOS (Farmer & Norton 1989; Abrams et al. 1996) and ACE (Hase et al. 2010). Their spectral coverage is similar (ATMOS: 625 – 4800 cm$^{-1}$; ACE: 750 – 4400 cm$^{-1}$). The ACE spectrum generally achieved a higher S/N, but at a lower resolution, 0.02 cm$^{-1}$ for ACE versus 0.01 cm$^{-1}$ for ATMOS. Both observations recorded the center of the solar disk with an aperture 10 – 15% of the solar diameter. Many unidentified lines appear in both: Hase et al. report in their reference [2] that 2400 newly identified lines in the ACE spectrum were published in 1995, and an additional unpublished 2700 lines in 1998.

The ACE spectrum was downloaded from its mission website, and included the plot of the atlas, in which many lines identified by L. Wallace are labeled; the two files of identified and unidentified lines; and the ASCII digital data file. The ACE digital data file provides pairs of wavenumber versus normalized intensity suitable for direct comparison against calculated spectra. To bring its lines into coincidence with those of both ATMOS and the calculated spectrum, the intensity was adopted of the wavenumber 0.005 cm$^{-1}$ higher than the original ACE value. Reduction of the individual FTS ATMOS strips into such pairs was accomplished by Kurucz. The ATMOS spectra have higher S/N than the ACE spectra at their bluest wavelengths, but their S/N declines below that of ACE at wavelengths >2.5 µm. Wherever both spectra existed, both were frequently consulted to confirm line detections; the higher resolution of the ATMOS spectra was often critical in disentangling line blends.

At shorter wavelengths, ground-based solar intensity spectra of the center of the solar disk were downloaded from the National Solar Observatory (NSO) website of Kitt Peak National Observatory (KPNO). The atlas of Wallace et al. (1993) was adopted over 730 – 11230 Å, and that of Wallace & Livingston (2003) from 11230 to 20830 Å. For each atlas, telluric spectrum removal was accomplished for all but the strongest line regions, based on a sequence of solar spectra obtained over a wide range of air mass.

Wallace et al. (2011) have provided a similarly high-resolution, high-S/N solar flux spectrum that includes light from the entire disk of the Sun, from which the telluric spectrum is also removed. They note that the solar flux spectrum is better suited as a standard for comparison against stellar spectra, which are necessarily flux spectra since stars are not spatially resolved. However, Figure 1 of Wallace et al. (2011) reveals that Fe I line profiles at disk center are sharper and deeper than those of the flux spectrum, and thus better suited to the present purpose. Unlike the flux spectrum, that of the disk center is free from the degradation in resolution engendered by the solar rotational velocity of 2 km s$^{-1}$. Moreover, the disk-center spectrum is obtained at an angle perpendicular to the solar photosphere, providing deeper penetration into the hotter solar layers that favor high-excitation Fe I line formation.

For Arcturus, two ground-based spectra were adopted: the Hinkle et al. (2000) atlas over 7300 – 9120 Å, and the Hinkle et al. (1995) atlas from 9100 Å to 5.3 µm. These were obtained with the FTS on the KPNO 4m telescope. Telluric removal for the Hinkle et al. (2000) atlas relied on the procedure and spectra noted above for the Sun. In the IR, Section 2.2 of Hinkle et al. (1995) describes how the Arcturus observations were scheduled to optimize telluric removal by maximizing the difference in the stellar geocentric velocity. The resulting summer and winter Arcturus spectra achieved displacements of at least 40 km s$^{-1}$, enough to readily distinguish individual absorption lines in Arcturus from those of the telluric spectrum.

The solar and Arcturus ground-based spectra were converted to wavenumber-intensity/flux pairs suitable for plotting using the data-reduction package IRAF. The tilt of each segment was removed to allow the end of one segment to agree with the beginning of the next, and the individual points in the overlap region were averaged. Regions or points that had experienced significant telluric contamination and showed extreme variations were removed. The two summer and winter spectra of Hinkle et al. (1995) were generally displayed individually at IR wavelengths. At wavelengths < 11100 Å, the summer spectrum alone was adopted, and was overplotted on the Hinkle et al. (2000) atlas for wavelengths where both were available.

## 6. INFRARED SPECTRAL SYNTHESIS

### 6.1 Calculations of Stellar Spectra in the Infrared

The spectral calculations followed the previous procedure, outlined in Section 5 of PK15 and further characterized in Section 3 of PKA17. Briefly, a one-dimensional (1D) radiative model atmosphere calculated in local thermodynamical equilibrium (LTE) of suitable temperature $T_{eff}$, gravity log g, and microturbulent velocity $v_t$ for each star was selected from the Castelli & Kurucz (2003) ODFNEW grid. The solar model was that of PK15, whose Table 1 lists its parameters. The models for the two cool giants Arcturus and HD 95870 were those described in Sections 3 and 5 of PKA17.

The SYNTHE program starts the spectral computation by calculating the radiative transfer through each of 17 rays of varying disk angle, from center to limb. For normal stars, SYNTHE then combines each ray with an appropriate offset dictated by the input stellar rotational velocity. This procedure, with zero rotation, was followed for the Arcturus spectral calculation. For the solar intensity spectral calculation, just the central ray calculation was retained. The microturbulent velocity $v_t$ = 1.0 km s$^{-1}$ adopted for the solar flux spectrum was reduced to 0.5 km s$^{-1}$ for the disk-center spectrum.

Each model spectrum was then generated by incorporating into the Kurucz SYNTHE program a list of wavelengths and transition probabilities for atomic and molecular transitions that give rise to absorption features observed in spectra of solar-temperature dwarfs and cooler giants. Calculations were performed in wavelength space, to conform to the adopted line lists, but were converted to wavenumber space in the search plots, in which an energy offset is a linear shift. The spectral calculations were run from 7300 Å to 6 μm (13,700 cm$^{-1}$ to 1680 cm$^{-1}$) for each star, and broadened by a Gaussian to match the observed breadth of the spectral line profile in each region for each star. Plots for the search were generated by overplotting the observed spectra described above on these calculations. Figure 1 of PKA17 provides an early example. From plots such as these, matches were sought to unidentified lines.

Spectral synthesis for this work in the IR was thus extended beyond the limit of 8900 Å of PK15, and the exploratory calculations near 18000 Å shown in Figure 1 of PKA17. To ensure representation of very weak lines in the IR calculation, the initial input atomic line list for elements with atomic number $Z \leq 30$ relied almost exclusively on the Kurucz (2018) quantum-mechanical calculations of lines of a given species for which both energy levels are known. In particular, this list for Fe I includes numerous, very weak lines arising from jK levels. Many of the published Fe I jK lines, for example those in Tables 2 and 3 of Johansson et al. (1994) but not those in Tables 3 and 4 of Schoenfeld et al. (1995), are available from the searchable database accessible through the NIST website. This is the primary source of atomic line parameters for many other theoretical spectral calculations, notably the Smith et al. (2021) update DR16 to the line list covering 15000 – 17000 Å (6670 – 5900 cm$^{-1}$) developed for the Apache Point Observatory Galactic Evolution Experiment (APOGEE; Majewski et al. 2017). Adopting the Kurucz lists assures that the jK lists are complete, and that all Fe I lines are on the same wavelength scale. It also provides calculated logarithmic gf values for every transition, to which a correction dgf can be derived to match the strength of absorption features discerned in stellar spectra.

The atomic line lists were downloaded from the Kurucz (1993) website from the subdirectory atoms under the subdirectory for each species, designated gf##*nn*.pos: ## = *Z,* and *nn* is the species ionization level, being 00 or 01 (neutral or singly ionized) for the cool stars considered here. Where available, lists were ingested that begin with 'hyper' as these include hyperfine splitting, which becomes increasingly important in the IR. For several species, lists were adopted that include isotopic splitting. Also adopted were lists that substitute laboratory gf values where they are available and appear to be improvements. Specifically, for Mg I, Mg II, Al I, Al II, P I, KI, Cr I, and Cr II, gf##*nn*.all were downloaded, plus gf1600.allauto for S I. gf##*nn*.alliso was adopted for Ti I and Ti II, while for Ni I, gf2800.posiso was adopted. For Na I, gf1100.all was updated from a download of gfall08oct17.dat from the Kurucz linelists/gfnew subdirectory. Since gfall08oct17.dat again includes hyperfine splitting for many species, it was also adopted for elements beyond $Z = 30$. For Ce II, the Cunha et al. (2017) list of nine H-band lines was incorporated. For Ba II, the Peterson et al. (2020) list developed by M. Spite incorporating isotopic splitting was adopted. Extensive revision to the calculated atomic gf values followed, to match the individual line strengths in each of the IR stellar spectra adopted. This work is still underway.

### *6.2 Validity of the Spectral Modeling in the Infrared*

Figure 1 plots (in wavenumbers) the observed spectra of the Sun and Arcturus superimposed on an earlier version of the IR calculations described in Section 5 of PKA17. These calculations for each star are shown as blue lines. The solar ACE spectrum in the 3.5µm region is plotted as a green line. The Hinkle et al. (1995) Arcturus observations in summer and winter are plotted as purple and orange lines; these are displaced downwards in *Y* by 0.1 for clarity. Along the top of the plot, the positions of selected lines of molecular and neutral atomic species are indicated by × for Fe I and by color-coded + signs for other species. The diamonds beneath them highlight Fe I lines that arise from jK levels.

The ACE solar spectrum shown in Figure 1 is of extremely high S/N. This proved essential to provide the large number of detectable lines needed to establish energies for 60 previously unidentified Fe I jK levels. A half-dozen jK lines fall within the 3.515 – 3.49 µm wavelength region plotted in Figure 1; they all arise from the s5f jK group of levels. One such line, at 2849.48 cm$^{-1}$, is from a level identified by Nave et al. (1994; hereafter N+94). The other five s5f lines are new identifications.

Regardless of their origin and identification status, all the absorption features in the solar spectrum over the region shown are weak. The *Y*-axis scale indicates that only the top 10% of the total intensity scale of the normalized solar spectrum is displayed. The calculations adopted Kurucz line lists that were generated prior to refined gf value adjustment, so the synthetic spectrum is too strong for many lines, such as those between 2844 – 2848 cm$^{-1}$. Unidentified lines appear most clearly in the solar spectrum; they are revealed as absorption features seen in the green (observed) spectrum that are deeper than in the blue (calculated) spectrum. All but one of the blue diamonds appear directly above such features, which were not included in these early calculations. Since the s5f line at 2849.48 cm$^{-1}$ was previously identified by N+94, this feature is matched by the calculation, and also has an × above it.

Figure 1 also indicates the significant extent to which line blending impacts the profiles of both known and unknown lines. It illustrates that this blending must be modeled accurately to reliably detect unidentified lines and determine their position and strength.

### *6.3 Line-Strength Variations in Infrared Spectra*

The Arcturus IR spectrum is generally too noisy to show the weakest lines of Fe I. Moreover, systematic differences are seen between its summer and winter spectra, for example at 2844.16 cm$^{-1}$ and 2847.48 cm$^{-1}$, where downward spikes appear in one Arcturus spectrum but not the other and no absorption line is seen in the Sun. Nonetheless the Arcturus spectra are useful to assess line blending.

Figure 1(b) of PKA17 illustrates that, while the Fe I line at 1776.5234 nm is about the same strength in the spectra of the Sun and Arcturus, the Mg I lines near 1774.96, 1775.4, and 1776.2 nm all become much stronger in Arcturus than in the Sun. This arises in part because Arcturus abundances of Mg and other light elements are enhanced with respect to iron (e.g., Peterson et al. 1993). Farther to the IR, Figure 1 indicates that the newly identified Fe I lines at 2844.68 cm$^{-1}$ and 2856.59 cm$^{-1}$ are mildly stronger in Arcturus than in the Sun. However, the line near the s5f Fe I line at 2850.19 cm$^{-1}$ is much stronger, in both the summer and winter spectra. This and the slight offset of the s5f blue diamond from the observed line suggest the latter is largely due to a different species, one whose lines grow in strength in Arcturus, such as an Mg I line whose gf value is too low. Any jK line as badly obscured as this one is not counted as detected in the search.

Figure 1 indicates that molecular lines are also stronger in Arcturus. CH lines are more than twice as strong in Arcturus than in the Sun, while the OH molecular lines are nearly 20 times stronger. In the solar spectrum these CH and OH lines are of the same strength as many unidentified lines, but the Arcturus spectrum ensures they will not be mistaken for them.

The calculations bear out the somewhat surprising result that the strength of many unidentified Fe I lines are at least as strong in the Arcturus IR spectrum as in the Sun (Figure 1; PKA17, Figure 1). This undoubtedly occurs because the continuous opacity beyond ~1.6 µm is much higher in the Sun than in a cool giant such as Arcturus, whose atmosphere is of low density. According to Figures 8.3 and 8.5(a) and (b) of Gray (2005), the IR continuum near 1.6 µm in both stars is formed at deep atmospheric levels because the H$^-$ bound-free opacity, the major source of continuous opacity, has declined dramatically from its maximum at 8500 Å. Beyond 1.6 µm in a warm star like the Sun, but not in a cool giant like Arcturus, the H$^-$ free-free opacity dominates, and rises steadily through the IR. In the Sun, by 2.0 µm the total continuous opacity of the Sun again equals that at 5000 Å, i.e. at a depth where $\tau_{5000} = 1$, the benchmark depth for optical continuum formation. Because the H$^-$ free-free opacity continues to rise at longer wavelengths, it drives solar line formation toward shallower, cooler levels.

In cool giants such as Arcturus, however, the low electron pressure greatly reduces the H$^-$ free-free opacity, and lines continue to be formed at large depths. Greater depths have higher temperatures, which leads to stronger lines of highly excited lines than expected for its effective temperature. This in turn suggests that highly excited Fe I lines that are detectable in the solar spectrum should remain as strong in giants that are cooler and more luminous. This is indeed observed, as discussed in Section 9.

Molecular lines also have a significant presence in the IR spectra of cool stars (e.g., Hinkle et al. 1995, Table 5). The line list adopted for molecular species is detailed in Section 5 of PKA17. These incorporate the gf value corrections by band, rather than to individual lines. Disagreements between calculated and observed molecular-line strengths were modest, with the exception of CO lines.

Disagreements for IR CO lines in the solar spectrum have been noted earlier, e.g., by Ayres & Testerman (1981) for the bands at 2200 and 4300 cm$^{-1}$. Both this work and Gray (2005) attributed the discrepancies to the enhanced formation of CO at low temperatures, coupled with the coexistence of hot and cold solar structures situated adjacent to one another at high atmospheric levels in the Sun. Indeed, images of the solar surface have long shown such structures as solar granulation (e.g., Title et al. 1989). Cheung et al. (2007) confirmed that these are due to warm rising upflows, which in the Sun turn over at $\tau_{5000} = 0.3$ (a photospheric height 130 – 140 km above $\tau_{5000} = 1$), and, at the granulation boundaries, descend as cooler flows at higher speed to large depths. For Arcturus, the strengths of IR CO lines, both at 4.7µm (e.g., Heasley et al. 1978; Wiedemann et al. 1994) and at 2.3 µm (Ohnaka & Morales Marín 2018), also suggest one or more layers of cool gas at high atmospheric levels in inhomogeneous structures. Modeling these structures is beyond the scope of this investigation.

## 7. INFRARED SEARCH PROCEDURES

From Figure 1, it is evident that the unidentified Fe I jK lines are usually found in blends with other features. The gf values of even very weak identified lines must be modified to assess the position and contribution of the identified feature to the blend. To do this across the entire span of the solar observations to the accuracy needed to match every known atomic transition is a daunting task. Furthermore, for CO it may be impossible to reach agreement using the 1D models adopted, due to the effect noted above of the spatial inhomogeneity of a convective atmosphere on CO line formation.

This extensive mismatch between calculated and observed spectra prevented the adoption of cross-correlation techniques to search for matched unidentified lines in this work. Cross-correlation techniques developed by Tonry & Davis (1979) have proven very effective in deriving radial velocities from spectral line shifts of individual spectra of stars in globular clusters (e.g., Peterson & Latham 1989; Peterson, et al. 1995). Here, however, the persistent, systematic nature of the mismatch has undesirable consequences. For example, each line whose strength is underestimated in the calculation generates a potential unidentified line to be matched. Masking individual lines from the cross-correlation is possible in principle, but infeasible with CO lines due to their high density and broad wavelength range (Hinkle et al. 1995, Table 5). Pending better modeling of the atomic and CO line strengths, the searches were conducted manually as before, following the earlier procedures.

Identifying jK levels is greatly simplified by the pattern of energies of levels within a particular jK group. Johansson et al. (1994) discussed this pattern as part of their determinations for levels of the s5g group of jK levels. Their Figure 2 shows that the energies of s5g levels fall within five distinct bands of width 5 – 15 cm$^{-1}$ separated by 100 – 400 cm$^{-1}$. The individual energies of the levels in each band occur in pairs whose angular momentum value J differs by 1. The mean energies of these pairs closely follow parabolas whose independent variable is constructed from j, K, and the angular momentum $l$ of the orbiting electron. Armed with this prescription, they reproduced the line positions observed in laboratory spectra and in the ATMOS solar atlas to derive energies good to 0.01 cm$^{-1}$ or better for 56 of the 58 levels present in the diagram. The two that escaped detection have the highest energies and the lowest J values. The scatter of the determined mean energies around each parabola is 0.02 cm$^{-1}$.

For this work we have exploited these relationships to determine energies for 60 unidentified levels in three jK configurations, those with core structure 4f, s5f, and s6f. Within each group of the same configuration, the total set of levels, known and unknown, was searched as a group. For the s6f group with no previous identifications, the first search was conducted for the energy of a level of high J in the lowest band. The spacing of two strong lines of this level with a modest difference in predicted wavenumbers was compared against observed spectra by finding two unidentified features with the identical spacing. Their mutual offset determined the shift of the predicted energy of that level to its actual energy, which was confirmed from its other detectable lines. This constrained the energies of each of the other levels of this band to typically ± 4 cm$^{-1}$. Searches within such a narrow range proved successful after a minimum of trial and error. The search then proceeded to bands of successively higher energy, by again choosing a high-J level and initially assuming the same offset as that found for the band beneath. Once its own detectable lines accurately specified the offset of the targeted level, other levels in the band were again found within a narrow energy range.

Attention was also paid to the number of strong lines of a level for which no line was detected at a level five times weaker than the predicted line strength. Levels were accepted only if the number of line detections exceeded by two or more the number of such strong nondetections. This constraint was adopted because the gf values of IR lines of jK levels are generally predicted quite accurately. Like the total ensemble of lines with dgf values (Section 8), they typically required an overall reduction of −0.1 or −0.2 dex, and the relative line gf values proved good to ±0.3 dex. Specifically, the dgf values of the 137 detected s6f lines with wavelengths > 900 Å in Table 3 average −0.094 dex, with a standard deviation of 0.334 dex. Those of the 258 detected s5f lines > 900 Å average −0.144 dex, with a standard deviation of 0.303 dex. Among both sets of levels, only 11 of these 395 IR lines have dgf values < −0.7 dex.

A new jK level energy was accepted if it fell within 1 cm$^{-1}$ of the parabola of its band and led to three or more line matches. In some cases, gf values of lines with wavelengths >6 μm, the limit of the spectral synthesis calculations, were high enough to suggest detectable lines in the ACE spectrum, as elaborated in Section 8 below. Consequently, successful matches to the digital wavelengths of relatively isolated features in the ACE plot were accepted, as well. As additional levels became established, the parabola was recalculated, and the conformity of the energies of less-certain individual levels was reexamined.

Among the jK levels evaluated in the three 4f, s5f, and s6f groups were the 45 levels identified by N+94. Of the 45, only the seven levels listed in Table 1 failed to meet these criteria. Two were reconciled by a slight change in the energy, and one by adopting a significant energy change. Energies of the other four were dropped from the parabola fits and the quantum-mechanical calculations, as they gave rise to a large number of strong lines at positions where none was observed. Energies of several s6d levels of even parity were similarly modified or dropped, and will be reported subsequently.

## 8. UNCERTAINTIES

Both LS and jK searches lead to uncertainties in the deduced energies that are significantly larger than most laboratory measurements. This is due primarily to the lack of IR lines in the LS levels, and to the weakness and blending of the jK lines in the stellar spectra. LS searches followed those of PK15 and PKA17; the latter work estimated an uncertainty of 0.01 cm$^{-1}$ at best, reaching 0.03 cm$^{-1}$ for levels with few lines or whose lines were restricted to the UV. The LS levels identified here should have similar uncertainties.

Compared to searches for LS levels, the jK searches are generally successful in shorter times, because the search for the energy of each jK level is limited to a much narrower range. Ambiguities do remain, however, especially when the width of a given band is narrow. Lines of similar energy and J overlap in wavelength and strength. The right energy may be assigned to the wrong level, and the resulting discrepancies may remain hidden due to low S/N, extreme line crowding, or telluric obliteration of critical-wavelength regions.

Energy uncertainties were estimated quantitatively from the shift between the calculated versus the observed position of a relatively unblended, newly matched Fe I line. This is best discerned at wavenumbers $<\sim 4760$ cm$^{-1}$ (wavelengths $>2.1$ μm), where the high-quality space-based spectra from ATMOS and ACE are available. A quantitative estimate was formed by comparing the wavenumber of the centroid in the digital ACE atlas of the profiles of the strongest IR lines against their calculated positions. The positional shift of a given line is the sum of the shifts in the lower and the upper energies. The transitions between s and s5f levels were the ones considered, since Schoenfeld et al. (1995) list errors to three decimal places for all the s6g levels. These transitions all lie at wavelengths $>7$ μm, and so have narrow profiles (Section 4), but no spectral synthesis is available as yet to identify blends and determine dgf values. Instead, all lines with a calculated log gf value $\geq +0.15$ dex were initially included. Of the 39 such lines, 17 lines were removed for which a blending feature was identified in the ACE atlas plot, and two more were dropped whose profiles were broad or asymmetric. The 20 remaining lines yielded an average shift of $-0.005$ cm$^{-1}$, with a standard deviation of $\pm 0.037$ cm$^{-1}$. This shift is consistent with the adoption for the spectral synthesis (Section 5) of the ACE intensity of the wavenumber 0.005 cm$^{-1}$ higher than the original ACE value, to bring the absorption features of the ACE spectrum into visual agreement with those of ATMOS and Arcturus. The mean of the Schoenfeld et al. (1995) errors for these lines is 0.021 cm$^{-1}$ and their standard deviation is $\pm 0.005$ cm$^{-1}$. Assuming that the errors are equal in the two sets of levels, the joint standard deviation of $\pm 0.037$ cm$^{-1}$ leads to an uncertainty of $\pm 0.026$ cm$^{-1}$ in the line positions of each group.

This is the standard deviation of an individual line position. As Schoenfeld et al. (1995) noted, the uncertainty of the energy of a given level depends strongly on the number of lines detected for that level and the blending of their profiles. Most IR jK lines are not as strong as those considered just above, and may be heavily blended in regions where lines arise from a multitude of jK levels (see below). However, Section 7 notes that for each level, three or more detections of jK lines and conformity to 1 cm$^{-1}$ of the parabola of its band were required to accept the level as identified. All of the s5f levels and most of the s6f levels have at least three IR line detections, and so should be good to $\pm 0.015$ cm$^{-1}$. The four s6f levels 7.0 s6f 4+[7+], 5.0 s6f 3+[4+], 2.0 s6f 4+[1+], and 1.0 s6f 3+[0+] have but two IR line detections, implying uncertainties of $\pm 0.020$ cm$^{-1}$. The 1.0 s6f 2+[1+] level has a single IR line detection. The detected line is distinct and largely unblended, yielding an uncertainty of $\pm 0.026$ cm$^{-1}$.

Errors in the calculated gf values were evaluated by forming a histogram of the derived dgf values. The 2100+ lines for which dgf values are listed in Table 3 yielded a mean dgf value of $-0.15$ dex, and a standard deviation of 0.38 dex. Deviations of 1.0 dex or more do occur but are rare: among the 2100+ lines with dgf values, 20 dgf values fell below $-1.15$ dex, and 27 were found above 0.85 dex. Except for the modest excess of these large deviations, the distribution of dgf values agrees well with a normal distribution.

Note that this work does not yet invoke the increase in damping constants that Schoenfeld et al. (1995) found necessary to match the broad wings of many lines of s6g levels. This will clearly be necessary here, as many s5f and s6f lines also show broad wings. Where the line profile was sufficiently unblended that such wings were revealed, some allowance for this was made in dgf by allowing the center of the line profile so calculated to be deeper than observed.

Whether a dgf value is truly representative of the amount of line absorption by the associated transition is most difficult to assess when the absorption on either side is poorly characterized. Many IR lines with dgf values were accepted simply to reduce the amount of unidentified absorption in the vicinity, and so to reduce systematic errors in future investigations that rely on continuum placement or flux distributions. The best method to confirm that a particular line absorbs at its precise wavelength is to check a high-quality solar atlas for an unmodeled depression that lies within ±0.02 cm$^{-1}$ of the line position. In the IR at wavenumbers < 3800 cm$^{-1}$, this is straightforward using the ACE atlas, from either the plot (which provides identification of blends), or the digital version (which provides precise positions). However, uncertainties become significantly larger at wavenumbers between 2340 – 2800 cm$^{-1}$, where absorption due to Fe I lines that remain unidentified is greater than the absorption by identified Fe I lines. This excess is evident from the number and strength of both the predicted Fe I lines and the unmatched features observed in the solar spectrum. Values of dgf in this region are rough approximations in most cases. Nonetheless, these lines, and the predicted lines without dgf values, should be included in calculations where modeling the continuum definition or flux distribution is important.

## 9. RESULTS

Applying the above procedures to UV, optical, and IR spectra has resulted in new identifications for nearly a hundred Fe I levels: one LS level of even parity, 38 LS levels of odd parity, and 60 jK levels of odd parity. Table 2 presents the full label of each newly identified level, its conventional shortened form, the value of its angular momentum J, and its energy E in cm$^{-1}$. LS levels precede jK levels. Within each group, levels are listed in order of increasing J, then increasing E. For jK levels, the labels give the full and abbreviated base configuration of the core followed by its values of j and K, the latter in square brackets. Each of these values is indicated by an integer and a + sign, which stands for .5.

An estimate of the number of detectable IR lines these levels may harbor can be derived following the gf value correction approach of PKA17. Just as the gf values of LS levels were corrected to the furnace temperature to account for the depletion of the population of the lower level of a transition as a function of its lower excitation energy $E_{lo}$, a similar corrected log $S$ = log gf –0.877×$E_{lo}$ can be invoked to account for the population at the solar photospheric temperature. Based on PKA17 and the new jK level detections, we count an IR line as detectable if log $S$ > –7.3 for wavelengths from 1.12 to 2.26 μm, –7.5 over 2.26 – 5 μm, –7.0 over 5 – 8 μm, –6.5 over 8 – 11 μm, and –5.5 over 11 – 16 μm. Applying these criteria to the list presented here of newly identified Fe I lines yields identifications for 143, 260, 201, 1, and 0 potentially detectable Fe I lines in each of these regions. This may be compared to the results of applying these criteria to the list of previously known Fe I lines, which results in 2606, 1968, 252, 64, and 8 known, potentially detectable Fe I lines in each of these regions. Among the previously known lines, jK levels accounted for 803, 508, 127, 1, and 1 lines. The jK level determinations are clearly very important in the 5 – 8 μm region, and have a significant presence over 2 – 5 μm as well.

## 10. STATUS AND POTENTIAL IMPACT OF Fe I LINE IDENTIFICATIONS

Table 3 provides a list of potentially detectable Fe I lines newly identified from the levels of Table 2, and their parameters and energy levels. All lines with log gf < –8 were removed, as were lines with log gf values below –7 > 300nm, –6 > 500 nm, –5 > 700 nm, –4 > 900 nm, –3 > 1000 nm, –2 >6000 nm, and  –1 >10000 nm, leaving over 16,000 Fe I lines from 157 nm to 16 μm. It is available online as an

.mrt file associated with this paper. It is also posted as table3.txt on the Kurucz website in subdirectory/atoms/2600/PK21. Its Readme file gives its format. Note that the value of log gf that best matches the line strength in stellar spectra is log(gf)o = log gf + dgf. Table 3 in the printed version lists just the first 12 lines. The untrimmed lists of all Fe I lines at all wavelengths that originate from these levels, plus those of the revised levels in Table 1 and all other identified Fe I lines, are likewise posted on the Kurucz website in subdirectory/atoms/2600. The file gf2600.lines includes lines with predicted energies, whose wavelengths are in error (Section 2). The file gf2600.pos provides only those lines of levels whose energies are both known (dropped levels excepted), and thus are suitable for high-resolution analysis.

Section 2 has outlined several areas of astrophysical research where the identification of numerous Fe I lines and levels can assist ongoing investigations. Here we describe the practical contributions of current and future identifications in three such areas: the modeling of UV fluxes, the reduction of systematic errors in the abundances of elements other than iron for warm stars of old stellar populations, and the determinations of abundances and abundance gradients for iron and the light elements from IR spectra of cool, luminous giants.

Figure 1 of PK15 illustrates that before our identifications of Fe I levels began, spectral calculations seriously underestimated the spectral fluxes in the 2550 – 2700 Å region of all but the most metal-poor stars of solar temperature. As Peterson et al. (2020) demonstrated, this deficiency is directly attributable to unidentified Fe I lines arising from levels of high excitation. The current work on LS levels has helped to close the gap for warm, metal-poor stars. However, the full benefit of such theoretical modeling is attained only when it reproduces the UV fluxes of the warm stars of solar metallicity. These are the stars that contribute the most UV flux in old solar-metallicity populations, which lack the hot blue horizontal branch stars of old, metal-poor populations. Their UV fluxes are thus sensitive to both metallicity and age: they are depressed as the iron content rises and absorbs more UV flux, but this also happens as old stellar systems age and the brightest warm stars become cool giants.

Modeling UV spectra of stars reliably is thus essential to derive age and metallicity determinations for old, quiescent galactic systems that are free from the long-standing ambiguity between age and metallicity (Worthey 1994). Some work has ventured toward incorporating the UV, but even the narrow-band Sloan *u* filter and its variants are centered near 350nm, where Fe I lines are outnumbered by lines of a variety of other species. Bringing UV modeling up to speed would take advantage of the specific sensitivity of the 2550 – 2700 Å region to the iron abundance, sharpening an essential tool for interpreting UV flux distributions observed at low resolution for many nearby galaxies. It could prove especially useful for disentangling age from metallicity and light-element enhancement in future James Webb Space Telescope (JWST) high-redshift targets, where the UV is shifted into the IR.

In future work we hope to successfully identify many of the high-J levels that are responsible for the strongest unidentified lines in the UV and optical. This is now possible especially for levels with J = 5 or higher, as the 2J+1 dependence of the statistical weight *g* (e.g., Eriksson & Lennerstad 2017) leads to higher gf values and thus stronger lines. The Kurucz calculations suggest that these levels not only have significant UV absorption, but also should give rise to strong and isolated unidentified lines in the IR. Indeed, the current work has failed to identify a good many of the strongest unidentified IR lines, notably those with wavelengths in the 1 – 2 μm range (e.g., Smith et al. 2021). The high-J states of Fe I occur in nf, ng, nh, and ni levels, which were not targeted in previous investigations because their predicted energies were frequently in error by 100 cm$^{-1}$ or more. The new level identifications can now confine the search for additional high-J LS levels to a much narrower range. As an example, only one out of 18 identified J = 5 odd-parity levels with energies > 60,000 cm$^{-1}$ shows an offset > 100 cm$^{-1}$.

By incorporating IR spectra and targeting specific groups of jK lines, this work has contributed a significant number of Fe I identifications throughout the IR (Section 9). Notably, it has nearly doubled the total number of Fe I lines identified in the 5 – 8 µm region. However, swarms of unidentified Fe I lines remain throughout the IR. The 5f, s7f, and s7h jK groups of odd parity, and the 5g, s7g, and s7i jK groups of even parity, all have high-J levels whose bands overlap in energy with the jK energies identified here, and so can likely yield many more successful identifications. This is also a high priority for our future work, to reduce the significant contribution by currently unidentified lines in extensive IR regions (Section 8).

Several areas of investigation would benefit. Improvements in the modeling of continuum placement and line blending in high-quality IR spectra of stars like the Sun and red horizontal branch stars (e.g., Afşar et al. 2018, Figure 3), would reduce systematic errors in abundance derivations of elements with few UV or optical lines. These include heavier elements past the iron peak represented primarily by UV lines (e.g., Peterson 2011; Roederer et al. 2016), and those whose IR lines are weak but detectable (e.g., Cunha et al. 2017). Also included are lighter elements with sparse spectra such as fluorine (e.g., Jönsson et al. 2014; Pilachowski and Pace 2015; Abia et al. 2019) and phosphorus (e.g., Maas et al. 2019). Heavy and light elements elucidate nucleosynthesis processes and environments at early epochs (Sneden et al. 2008), and the delayed contribution of evolved asymptotic giant-branch stars (Pilachowski & Pace 2015). Phosphorus is of special importance to astrobiology, as its relative abundance in a planetary host star influences prospects for life on its planet (Hinkel et al. 2020).

Systematic errors in the IR analysis of the iron abundance of cooler, stronger-lined red giants are also reduced by identifications of IR Fe I lines. As discussed in Section 6.3, due to the reduction of the H$^-$ free-free opacity in the low densities of giant atmospheres, highly excited Fe I lines in IR spectra of giants are expected to have the same or greater strength than do those in the solar spectrum. Consequently, the numbers in Section 9 of newly-detectable IR lines in the Sun should also apply to spectra of cool but luminous giants. Observations confirm this: unidentified lines are widely present in IR spectra of cool, luminous giants, even the coolest of these. Johansson et al. (1991) noted them around 4 µm in the spectrum of the cool K5 giant α Tau (Ridgway et al. 1984). Figure 2 of García Pérez et al. (2015) reveals them as sharp peaks in the spectrum of the difference between the best spectral fit for a cool, solar-metallicity giant and its APOGEE spectrum. Figure 3 of Ryde et al. (2016) shows 2 µm spectra for luminous red giants in the inner Galactic bulge, with many lines unmatched by spectral synthesis in the strongest-lined star.

The identification of the jK levels in Table 2, and of additional groups of jK levels, should surely lead to more reliable IR continuum placement in cool giants. Defining the IR continuum is crucial where optical spectra are too faint to be obtained, such as in giants of the bulge, bar, and disk of the Milky Way (e.g., Ryde et al. 2016). This would provide more accurate abundances and abundance gradients of iron and the light-alpha elements, refining the understanding of the assembly and dissipation of these structures. Moreover, the improvement in estimates of reddening by Galactic dust (e.g., Schlafly et al. 2016) should assist interpretations of JWST bulge and distant IR sources.

We thank the anonymous referee for comments which helped to improve this paper. IRAF is distributed by the National Optical Astronomy Observatories (NOAO), which are operated by Association of Universities for Research in Astronomy, Inc. (AURA), under cooperative agreement with the National Science Foundation (NSF). The solar NSO/Kitt Peak FTS data were produced by NSF/NOAO. We thank Frank Hill for assistance with obtaining them. We thank M. Afşar and C. Sneden for the spectrum of HD 95870 (HIP 54048), which was obtained with the Immersion Grating

Infrared Spectrometer (IGRINS) that was developed under a collaboration between the University of Texas at Austin and the Korea Astronomy and Space Science Institute (KASI) with the financial support of NSF under grants AST1229522 and AST-1702667, of the University of Texas at Austin, and of the Korean GMT Project of KASI. Support for Fe I identifications under program number HST-GO-15179 was provided to the authors through National Aeronautics and Space Administration (NASA) grants from the Space Telescope Science Institute, which is operated by AURA under NASA contract NAS 5-26555. Support for the IR extension of Fe I identifications was provided by NASA grant No. 80NSSC19K0750. This research is based on UV spectra from the NASA/ESA Hubble Space Telescope (HST) from the following HST programs, whose PI is in parentheses: GO-7348 (Edvardsson), SNAP-7402 (Peterson), GO-8197(Duncan), GO-9146 (Evans), GO-9455 (Peterson), GO-9491 (Balachandran), GO-9804 (Duncan), GO-14161 (Peterson), GO-14762 (Peterson), and GO-15179 (Peterson). This work has also incorporated optical spectra, including those from European Southern Observatory programs 080.D-0347(A) and 082.B-610(A); from programs 2006-A-C-5 and 2006-B-Q-47 of the international Gemini Observatory, a program of NSF's NOIRLab, which is managed by AURA under a cooperative agreement with NSF; and from W. M Keck Observatory programs U17H, U35H, U44H, U63H, H6aH, N01, N12, and N13H, largely accessed through the Keck Observatory Archive (KOA), which is operated by the W. M. Keck Observatory and the NASA Exoplanet Science Institute (NExSci), under contract with NASA. Table 1 of PK15 provides details. This work has likewise depended on IR solar spectra from two space missions, ACE and ATMOS. The Atmospheric Chemistry Experiment (ACE), also known as SCISAT, is a Canadian-led mission mainly supported by the Canadian Space Agency. The Atmospheric Trace Molecule Spectroscopy Experiment (ATMOS) team research has been carried out at the Jet Propulsion Laboratory, California Institute of Technology, under contract to the National Aeronautics and Space Administration.

FIGURE 1

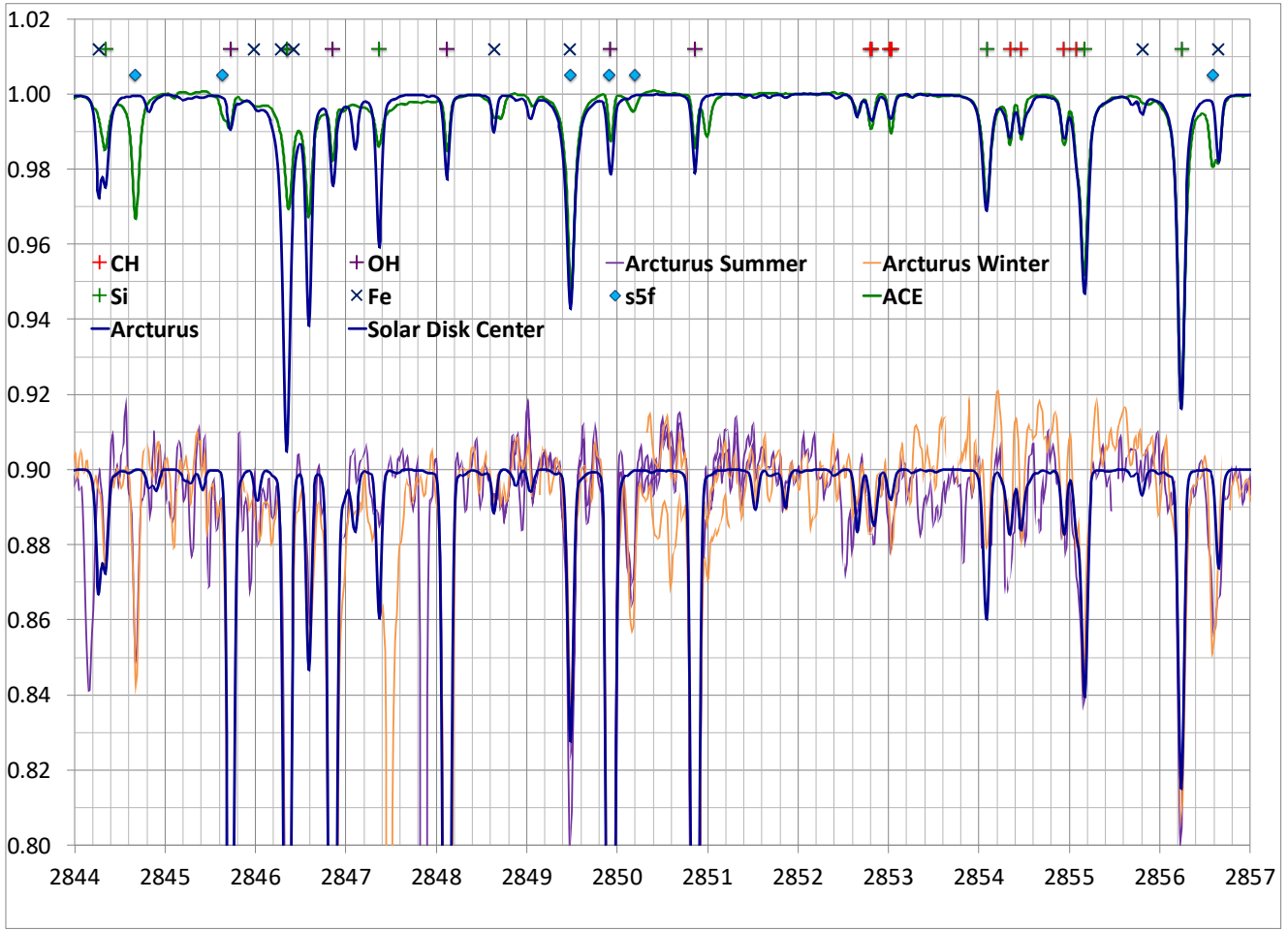

Figure 1. Observations vs. calculations for the continuum-normalized intensity of the Sun and of the flux of Arcturus over 3.515 – 3.49μ. On the *X*-axis appears the wavenumber in cm$^{-1}$. On the *Y*-axis is the intensity scale for the solar spectrum; the Arcturus spectrum is offset vertically by –0.1 for clarity. The observed spectrum for the Sun is that of ACE (Hase et al. 2010; green line). For Arcturus are shown both the winter (orange line) and summer (purple line) spectra of Hinkle et al. (1995). Superimposed as solid blue lines are spectral synthesis calculations (Section 6.2) for the center of the solar disk and the integrated flux of Arcturus. The positions of selected lines of molecular and neutral atomic species are indicated by × for Fe I and by color-coded + signs for the other species indicated. The diamonds beneath them highlight Fe I lines that arise from s5f jK levels. The s5f line at 2849.5 cm$^{-1}$ was previously identified by Nave et al. (1994). The remaining five s5f lines, which fall above observed features that are unmatched except for line blends, are new identifications.

TABLE 1

Revised Fe I Levels and Energies

| Expanded Label | Label | J | E (cm$^{-1}$) | Former E (cm$^{-1}$) | Reference |
|---|---|---|---|---|---|
| LS levels: | | | | | |
| *Five even levels:* | | | | | |
| 3d6 4s(6D)8s 5D | 4s6D8s 5D | 3 | Drop | 60078.83 | PKA17 |
| 3d7(2G)5s 3G | (2G)5s 3G | 3 | Drop | 61724.84 | PK15 |
| 3d6 4s(6D)6d 3+[4+] | s6d 3+[4+] | 4 | 59532.84 | 59532.97 | PK15 |
| 3d7(2G)5s 3G | (2G)5s 3G | 5 | 61198.48 | 61198.49 | PK15 |
| 3d7(2G)4d 3I | (2G)4d 3I | 6 | 67185.97 | 67245.33 | PKA17 |
| 3d7(2G)4d 3I | (2G)4d 3I | 7 | 67167.93 | 67167.95 | PKA17 |
| *One odd level:* | | | | | |
| d6(5D)4s(6D)7p 7F | 4s6D7p 7F | 5 | 59265.72 | 59254.76 | PKA17 |
| jK levels: | | | | | |
| *Seven odd levels*: | | | | | |
| d7(4D)4f 2+[2+] | 4f 2+[2+] | 2 | Drop | 59680.306 | N+94 |
| d7(4D)4f 4+[3+] | 4f 4+[3+] | 3 | Drop | 58714.648 | N+94 |
| d7(4D)4f 1+[4+] | 4f 1+[4+] | 4 | Drop | 59960.956 | N+94 |
| d7(4D)4f 4+[5+] | 4f 4+[5+] | 5 | 58700.81 | 58700.834 | N+94 |
| d7(4D)s5f 3+[5+] | s5f 3+[5+] | 5 | Drop | 59663.964 | N+94 |
| d7(4D)s5f 4+[6+] | s5f 4+[6+] | 6 | 59280.56 | 59275.884 | N+94 |
| d7(4D)s5f 3+[6+] | s5f 3+[6+] | 6 | 59670.06 | 59670.084 | N+94 |

TABLE 2

New Fe I Levels and Energies

| Expanded Label | Label | J | E (cm$^{-1}$) |
|---|---|---|---|
| LS levels: | | | |
| *One even level:* | | | |
| 3d7(4F)5d 5H | (4F)5d 5H | 5 | 59687.88 |
| *38 odd levels:* | | | |
| d6(5D)4s(6D)5p 5D | 4s6D5p 5D | 0 | 58556.38 |
| d7(4F)6p 5F | (4F)6p 5F | 1 | 59761.68 |
| d6(5D)4s(6D)7p 5P | 4s6D7p 5P | 1 | 60253.31 |
| d6(5D)4s(6D)8p 7D | 4s6D8p 7D | 1 | 61266.98 |
| d6(5D)4s(6D)8p 5P | 4s6D8p 5P | 1 | 61621.21 |
| d6(5D)4s(6D)9p 5D | 4s6D9p 5D | 1 | 62103.84 |
| d7(4F)7p 5F | (4F)7p 5F | 1 | 62417.13 |
| d7(4F)7p 3D | (4F)7p 3D | 1 | 62666.97 |
| d6(5D)4s(4D)5p 3F | 4s4D5p 3F | 2 | 58483.58 |
| d6(5D)4s(6D)7p 7D | 4s6D7p 7D | 2 | 59632.62 |
| d6(5D)4s(6D)7p 5F | 4s6D7p 5F | 2 | 60099.96 |
| d6(5D)4s(6D)8p 7D | 4s6D8p 7D | 2 | 60999.34 |
| 61023o | 61023o | 2 | 61023.23 |
| d6(5D)4s(6D)8p 5D | 4s6D8p 5D | 2 | 61208.17 |
| d6(5D)4s(6D)8p 7F | 4s6D8p 7F | 2 | 61285.50 |
| d6(5D)4s(6D)8p 5P | 4s6D8p 5P | 2 | 61432.02 |
| d6(1D)4s4p(3P) 3P | 1Dsp3P 3P | 2 | 61826.24 |
| d6(5D)4s(6D)9p 5P | 4s6D9p 5P | 2 | 62201.85 |
| d7(4F)7p 5F | (4F)7p 5F | 2 | 62273.34 |
| d6(5D)4s(6D)9p 7P | 4s6D9p 7P | 2 | 62326.92 |
| d7(4F)7p 3D | (4F)7p 3D | 2 | 62331.33 |
| d6(5D)4s(6D)10p 7D | 4s6D10p 7D | 2 | 62386.97 |
| d6(5D)4s(6D)7p 7D | 4s6D7p 7D | 3 | 59339.26 |
| d6(5D)4s(6D)7p 7F | 4s6D7p 7F | 3 | 59503.40 |
| d6(5D)4s(6D)8p 5D | 4s6D8p 5D | 3 | 61012.24 |
| d6(5D)4s(6D)9p 5P | 4s6D9p 5P | 3 | 61476.76 |
| d6(5D)4s(6D)9p 7F | 4s6D9p 7F | 3 | 62268.60 |
| d7(4F)6p 5F | (4F)6p 5F | 4 | 59117.74 |
| d6(5D)4s(6D)8p 5D | 4s6D8p 5D | 4 | 60628.77 |
| d6(5D)4s(6D)8p 7F | 4s6D8p 7F | 4 | 61173.80 |
| d7(4F)7p 5D | (4F)7p 5D | 4 | 61232.91 |
| d6(5D)4s(6D)10p 5F | 4s6D10p 5F | 4 | 62537.38 |
| d6(3G)4s4p(1P) 3H1 | 3Gsp1P 3H1 | 4 | 63501.17 |
| d6(3G)4s4p(1P) 3H2 | 3Gsp1P 3H2 | 4 | 63571.96 |
| d7(4F)7p 5G | (4F)7p 5G | 6 | 61173.01 |
| d6(3H)4s4p(3P) 5I | 3Hsp3P 5I | 7 | 42880.56 |
| d6(3H)4s4p(3P) 5H | 3Hsp3P 5H | 7 | 43348.46 |
| d6(3H)4s4p(3P) 5I | 3Hsp3P 5I | 8 | 43099.26 |

jK levels:

*60 odd levels:*

| | | | |
|---|---|---|---|
| d6(5D)4s(6D)4f 2+[0+] | s4f 2+[0+] | 0 | 57420.84 |
| d7(4F)4f 3+[1+] | 4f 3+[1+] | 1 | 59270.59 |
| d6(5D)4s(6D)5f 4+[1+] | s5f 4+[1+] | 1 | 59297.42 |
| d6(5D)4s(6D)6f 4+[1+] | s6f 4+[1+] | 1 | 60658.52 |
| d6(5D)4s(6D)6f 3+[0+] | s6f 3+[0+] | 1 | 61037.83 |
| d6(5D)4s(6D)6f 2+[1+] | s6f 2+[1+] | 1 | 61316.46 |
| d6(5D)4s(6D)5f 4+[1+] | s5f 4+[1+] | 2 | 59297.38 |
| d6(5D)4s(6D)5f 3+[2+] | s5f 3+[2+] | 2 | 59668.53 |
| d6(5D)4s(6D)5f 2+[2+] | s5f 2+[2+] | 2 | 59955.05 |
| d6(5D)4s(6D)5f 1+[1+] | s5f 1+[1+] | 2 | 60141.46 |
| d6(5D)4s(6D)5f 1+[2+] | s5f 1+[2+] | 2 | 60150.82 |
| d6(5D)4s(6D)6f 4+[2+] | s6f 4+[2+] | 2 | 60654.52 |
| d6(5D)4s(6D)6f 4+[1+] | s6f 4+[1+] | 2 | 60657.10 |
| d6(5D)4s(6D)6f 3+[2+] | s6f 3+[2+] | 2 | 61036.11 |
| d6(5D)4s(6D)6f 3+[1+] | s6f 3+[1+] | 2 | 61037.02 |
| d6(5D)4s(6D)6f 2+[2+] | s6f 2+[2+] | 2 | 61319.92 |
| d6(5D)4s(6D)6f 0+[2+] | s6f 0+[2+] | 2 | 61626.82 |
| d6(5D)4s(6D)5f 2+[3+] | s5f 2+[3+] | 3 | 59954.48 |
| d6(5D)4s(6D)5f 2+[2+] | s5f 2+[2+] | 3 | 59956.10 |
| d6(5D)4s(6D)5f 1+[2+] | s5f 1+[2+] | 3 | 60147.82 |
| d6(5D)4s(6D)5f 1+[3+] | s5f 1+[3+] | 3 | 60153.56 |
| d6(5D)4s(6D)5f 0+[2+] | s5f 0+[2+] | 3 | 60260.28 |
| d6(5D)4s(6D)5f 0+[3+] | s5f 0+[3+] | 3 | 60262.24 |
| d6(5D)4s(6D)6f 4+[3+] | s6f 4+[3+] | 3 | 60650.19 |
| d6(5D)4s(6D)6f 4+[2+] | s6f 4+[2+] | 3 | 60653.66 |
| d6(5D)4s(6D)6f 1+[3+] | s6f 1+[3+] | 3 | 61523.65 |
| d6(5D)4s(6D)6f 0+[3+] | s6f 0+[3+] | 3 | 61627.26 |
| d6(5D)4s(6D)5f 4+[3+] | s5f 4+[3+] | 4 | 59283.87 |
| d6(5D)4s(6D)5f 3+[3+] | s5f 3+[3+] | 4 | 59667.29 |
| d6(5D)4s(6D)5f 2+[3+] | s5f 2+[3+] | 4 | 59953.89 |
| d6(5D)4s(6D)5f 1+[4+] | s5f 1+[4+] | 4 | 60143.31 |
| d6(5D)4s(6D)5f 1+[3+] | s5f 1+[3+] | 4 | 60152.95 |
| d6(5D)4s(6D)5f 0+[3+] | s5f 0+[3+] | 4 | 60260.45 |
| d6(5D)4s(6D)6f 4+[4+] | s6f 4+[4+] | 4 | 60646.88 |
| d6(5D)4s(6D)6f 4+[3+] | s6f 4+[3+] | 4 | 60648.91 |
| d6(5D)4s(6D)6f 3+[3+] | s6f 3+[3+] | 4 | 61034.35 |
| d6(5D)4s(6D)6f 2+[4+] | s6f 2+[4+] | 4 | 61319.16 |
| d6(5D)4s(6D)6f 0+[3+] | s6f 0+[3+] | 4 | 61626.10 |
| d6(5D)4s(6D)5f 4+[5+] | s5f 4+[5+] | 5 | 59276.69 |
| d6(5D)4s(6D)5f 4+[4+] | s5f 4+[4+] | 5 | 59280.61 |
| d6(5D)4s(6D)5f 3+[4+] | s5f 3+[4+] | 5 | 59666.81 |
| d6(5D)4s(6D)5f 2+[5+] | s5f 2+[5+] | 5 | 59949.47 |
| d6(5D)4s(6D)5f 1+[4+] | s5f 1+[4+] | 5 | 60142.00 |
| d6(5D)4s(6D)6f 4+[5+] | s6f 4+[5+] | 5 | 60645.39 |
| d6(5D)4s(6D)6f 4+[4+] | s6f 4+[4+] | 5 | 60646.34 |
| d6(5D)4s(6D)6f 3+[4+] | s6f 3+[4+] | 5 | 61033.58 |
| d6(5D)4s(6D)6f 3+[5+] | s6f 3+[5+] | 5 | 61033.70 |
| d6(5D)4s(6D)6f 2+[5+] | s6f 2+[5+] | 5 | 61316.16 |
| d6(5D)4s(6D)6f 2+[4+] | s6f 2+[4+] | 5 | 61317.97 |
| d6(5D)4s(6D)5f 3+[5+] | s5f 3+[5+] | 6 | 59666.41 |
| d6(5D)4s(6D)5f 2+[5+] | s5f 2+[5+] | 6 | 59948.11 |
| d6(5D)4s(6D)6f 4+[5+] | s6f 4+[5+] | 6 | 60644.57 |
| d6(5D)4s(6D)6f 4+[6+] | s6f 4+[6+] | 6 | 60647.32 |
| d6(5D)4s(6D)6f 3+[5+] | s6f 3+[5+] | 6 | 61033.59 |
| d6(5D)4s(6D)6f 3+[6+] | s6f 3+[6+] | 6 | 61034.68 |
| d6(5D)4s(6D)6f 2+[5+] | s6f 2+[5+] | 6 | 61315.77 |
| d6(5D)4s(6D)6f 4+[6+] | s6f 4+[6+] | 7 | 60646.33 |
| d6(5D)4s(6D)6f 4+[7+] | s6f 4+[7+] | 7 | 60653.92 |
| d6(5D)4s(6D)6f 3+[6+] | s6f 3+[6+] | 7 | 61035.01 |
| d6(5D)4s(6D)6f 4+[7+] | s6f 4+[7+] | 8 | 60653.96 |

Note. The Kurucz website in the subdirectory/atoms/2600 contains the log files from the least-squares fits, b2600e.log and b2600o.log, which list the three largest eigenvector components for each level.

## TABLE 3

### Newly Classified Lines of Fe I

| Wavelength (nm) | log gf | <dgf | $E_e$ (cm$^{-1}$) | Je | Level_e | $E_{odd}$ (cm$^{-1}$) | J_o | Level_odd | $\Gamma_R$ | $\Gamma_S$ | $\Gamma_W$ | Ref |
|---|---|---|---|---|---|---|---|---|---|---|---|---|
| 157.7718 | -6.035 | ... | 0.000 | 4 | 4s2 a5D | 63382.670 | 5 | 3Gsp1P 3H | 8.65 | -3.77 | -7.09 | PKA17 |
| 158.2708 | -4.934 | ... | 0.000 | 4 | 4s2 a5D | 63182.860 | 4 | (4F)8p 3G | 7.99 | -3.41 | -6.94 | PKA17 |
| 158.3380 | -4.790 | ... | 415.933 | 3 | 4s2 a5D | 63571.960 | 4 | 3Gsp1P 3H2 | 8.54 | -3.90 | -7.12 | PKLS |
| 158.5157 | -4.813 | ... | 415.933 | 3 | 4s2 a5D | 63501.170 | 4 | 3Gsp1P 3H1 | 8.42 | -3.74 | -7.07 | PKLS |
| 159.0210 | -6.960 | ... | 0.000 | 4 | 4s2 a5D | 62884.790 | 3 | (4F)8p 3F | 8.25 | -3.06 | -7.02 | PKA17 |
| 159.3196 | -7.628 | ... | 415.933 | 3 | 4s2 a5D | 63182.860 | 4 | (4F)8p 3G | 7.99 | -3.41 | -6.94 | PKA17 |
| 159.3696 | -4.303 | ... | 0.000 | 4 | 4s2 a5D | 62747.240 | 4 | 4s6D10p 7F | 7.41 | -3.05 | -6.79 | PKA17 |
| 159.5309 | -5.035 | ... | 0.000 | 4 | 4s2 a5D | 62683.770 | 4 | 8p 3G5G5F | 8.20 | -3.21 | -7.01 | PK15 |
| 159.5522 | -4.783 | ... | 0.000 | 4 | 4s2 a5D | 62675.420 | 5 | (4F)8p 5G | 7.75 | -2.65 | -6.96 | PKA17 |
| 159.5946 | -2.934 | ... | 0.000 | 4 | 4s2 a5D | 62658.780 | 3 | 4s6D10p 5D | 7.99 | -3.33 | -6.82 | PKA17 |
| 159.8022 | -6.420 | ... | 0.000 | 4 | 4s2 a5D | 62577.370 | 4 | 8p 3G3F3G | 8.28 | -3.31 | -7.02 | PKA17 |
| 159.9044 | -3.100 | ... | 0.000 | 4 | 4s2 a5D | 62537.380 | 4 | 4s6D10p 5F | 7.86 | -3.41 | -6.82 | PKLS |

Note. Table 3 is published in its entirety in the electronic edition of *The Astrophysical Journal Supplement Series*. A portion is shown here for guidance regarding its form and content. The column marked by < before the dgf column contains the < character if the dgf value is an upper limit.

(This table is available in its entirety in machine-readable form.)